\documentclass[aps,prl,twocolumn, superscriptaddress,groupedaddress,nofootinbib,10pt,preprintnumbers]{revtex4-1}  
\usepackage{graphicx}  
\usepackage{dcolumn}   
\usepackage{bm}        
\usepackage{amssymb}   
\usepackage[caption=false]{subfig}
\usepackage{xcolor} 

\newcommand{\cO}{\mathcal{O}}
\newcommand{\gappeq}{\mathrel{\rlap {\raise.5ex\hbox{$>$}}
{\lower.5ex\hbox{$\sim$}}}}
\newcommand{\lappeq}{\mathrel{\rlap{\raise.5ex\hbox{$<$}}
{\lower.5ex\hbox{$\sim$}}}}

\begin{document}

\title{String Theory and the Dark Glueball Problem}
\author{James Halverson} \affiliation{Department of Physics, Northeastern University, Boston, MA 02115-5000 USA} 
\author{Brent D. Nelson} \affiliation{Department of Physics, Northeastern University, Boston, MA 02115-5000 USA} 
\author{Fabian Ruehle} \affiliation{Deutsches Elektronen-Synchrotron DESY, 22607 Hamburg, Germany}
\preprint{DESY-16-170}

\begin{abstract}
  We study cosmological constraints on dark pure Yang-Mills sectors.
  Dark glueballs are overproduced for large regions of ultraviolet
  parameter space.  The problem may be alleviated in two ways: via a
  large preferential reheating into the visible sector, motivating
  certain inflation or modulus decay models, or via decays into axions
  or moduli, which are strongly constrained by nucleosynthesis and  $\Delta N_{\text{eff}}$
  bounds.  String models frequently have multiple hidden Yang-Mills sectors,
  which are subject to even stronger constraints due to the existence of multiple dark glueballs.
\end{abstract}

\pacs{}
\maketitle

\noindent\textbf{Introduction.}

Hidden gauge sectors are ubiquitous in string theory.  Initially they
arose in the perturbative heterotic string, beginning with the
ten-dimensional $E_8\times E_8$ heterotic string \cite{Gross:1984dd}
itself, its orbifold compactifications
\cite{Dixon:1985jw,Dixon:1986jc,Ibanez:1986tp,Ibanez:1987pj,Lebedev:2006kn,Blaszczyk:2009in},
 free fermionic
realizations (e.g. \cite{Antoniadis:1989zy,Faraggi:1997dc}), and later
its smooth Calabi-Yau
\cite{Braun:2005ux,Bouchard:2005ag,Anderson:2012yf}
compactifications. They also arise in other contexts, for example on
$D$-branes in type II models (e.g. \cite{Cvetic:2004ui,Gmeiner:2005vz,Blumenhagen:2008zz}); in RCFT orientifolds
\cite{Dijkstra:2004ym,Dijkstra:2004cc}; on singularities in $G_2$
compactifications of M-theory (e.g. \cite{Acharya:1998pm,MR1424428,*Joyce2,Halverson:2015vta}); and on seven-branes in F-theory. In
fact, in F-theory there is growing evidence \cite{Grassi:2014zxa,Halverson:2015jua,Taylor:2015ppa,Taylor:2015xtz,Halverson:2016vwx} that dark gauge
sectors are generic in a sharp sense: the set of seven-brane
configurations at a generic point in seven-brane moduli space has
multiple disconnected gauge sectors.

In this paper we study the cosmological implications of gauged
hidden sectors from an ultraviolet (UV) perspective, focusing on the
simplest case of pure Yang-Mills theory. This is well motivated: if
$N$ of the hidden gauge factors in a much larger set of hidden sectors
are either pure super Yang-Mills or have no symmetry protection for
matter masses (other than supersymmetry), then $N$ pure Yang-Mills
sectors arise in the infrared. The associated $N$ UV gauge couplings
may each take a variety of values, determined by moduli stabilization,
giving $N$ hierarchical confinement scales due to the exponential
dependence of each on its UV gauge coupling.

These sectors are   cosmologically relevant if reheating via inflaton
or modulus decay reheats not only the visible sector, but some of the
hidden sectors as well. Such a scenario was studied in the case of a
single hidden sector in \cite{Carlson:1992fn}. Though not focused
specifically on dark glueballs, the relic abundance inferred from
\cite{Carlson:1992fn} depends critically on the confinement scale and
the ratio of visible to hidden sector entropy densities determined by
reheating; this is the result from which many of ours follow. This
scenario also has an effective $3\to 2$ self-interaction that causes
the dark sector to ``cannibalize'' itself, a phenomenon of recent
interest~\cite{Hochberg:2014dra,Pappadopulo:2016pkp,Forestell:2016qhc,Farina:2016llk}.
Dark matter that is comprised of dark glueballs has also been the
subject of a number of
studies~\cite{Faraggi:2000pv,Feng:2011ik,Boddy:2014yra,Soni:2016gzf,Kribs:2016cew}.

In this paper we will show the converse: in ultraviolet theories (such
as string theory) with many Yang-Mills sectors and a variety of dark
confinement scales, the associated dark glueballs are poor dark matter
candidates, but place valuable cosmological constraints on the
ultraviolet theory. The problem exists already in the case of a single
dark glueball, as its relic abundance oversaturates the observed dark
matter relic abundance for much of the natural UV parameter
space. This oversaturation is simple to understand, as the dark
confinement scale may take a variety of values and there is no reason
to have a dark glueball ``miracle'' analogous to the WIMP miracle. The
problem is exacerbated in theories with many dark glueballs since each
may have a different confinement scale, and if any falls into the
dangerous regions of parameter space that glueball will oversaturate.

We study two possible ways that the problem may be ameliorated, via
dark glueball decay into axions or moduli, or via preferential
reheating into the visible sector. Each mechanism faces constraints of
their own, the former from nucleosynthesis bounds on glueball lifetimes and on the
effective number of neutrinos $\Delta N_{\rm eff}$ present at late
times, and the latter on inflationary or modulus decay model building.
In the case of symmetric reheating most of the parameter space is ruled out,
even after taking into account possible decays.

\vspace{.5cm}\noindent\textbf{The Relic Abundance of Dark Glueballs.}

We consider a scenario in which a dark Yang-Mills sector with gauge
group $G$ and confinement scale $\Lambda$ is reheated to a temperature
$T_{\text{rh}}' > \Lambda$. The dark sector is a thermal bath of dark gluons,
and as the dark sector cools through a transition temperature
$T_\Lambda' \sim \Lambda$ the energy density in gluons is converted
into glueballs. For dark sector temperature $T' < T'_\Lambda$,
number depleting $3\to 2$ interactions change the dependence of
$T'$ on the scale factor $a(t)$ relative to that of non-interacting
non-relativistic particles, giving a dark to visible temperature ratio 
\begin{equation}
\frac{T'}{T} \propto \frac{a}{\ln(a)}\,.
\end{equation}
Physically, this unusual temperature dependence arises because the
interactions increase the average kinetic energy per glueball, so the
dark sector ``cannibalizes'' itself to stay warm.  Freezeout occurs
when these interactions cease to be effective, leaving a dark glueball
relic.

Through this process, comoving entropy density is conserved in each
sector due to thermal equilibrium and minimal interactions between the sectors, so that the
ratio
\begin{equation}
\xi := \frac{s}{s'}
\label{eq:xidef}
\end{equation}
is a constant. For sufficiently high $T_{\text{rh}}$ both sectors are
relativistic, since the dark sector is by the assumption
$T_{\text{rh}}'>\Lambda$, giving the additional relation $\xi = g_S T^3/ g_S'
T'^3$. In this case the initial entropy ratio could instead be thought
of as an initial temperature ratio, $\xi_T:= T/T' = (g_S'\, \xi /
g_S)^{1/3}$. We call $\xi=1$ the democratic scenario. 

\vspace{.5cm} This cosmological scenario was studied by Carlson et
al. in \cite{Carlson:1992fn}, which treated the lightest glueball as a
scalar field $\phi$. Let us review their results. The annihilation
rate for an average particle via $3\to 2$ interactions is determined
by an effective operator
\begin{equation}
\cO_{3\to 2} = \frac{1}{\Lambda} \frac{f}{5!} \phi^5\,,
\end{equation}
with the rate given by
\begin{equation}
\Gamma(3\to2) = \frac{\sqrt{5}f^2n'^2}{2304\pi\Lambda^5} = \lambda \Lambda \left(\frac{n'}{\Lambda^3} \right)^2\,,
\end{equation}
where $\lambda = \sqrt{5}f^2/(2304\pi)$. As the universe expands this
rate is eventually not high enough to further deplete the glueball
number and therefore they decouple at a temperature $T_d' \leq
\Lambda$. Using the fact that $\xi$ is constant and comparing to the
visible sector entropy density today, the relic abundance is
\begin{equation}
\Omega h^2 = \frac{T_d'}{3.6\, {\rm eV}\, \xi}\,.
\label{eq:relicvsTd}
\end{equation}
At decoupling, Einstein's equations may be radiation or matter
dominated. In the case of radiation domination at decoupling, $T_d'$
may be determined by solving the transcendental equation (the small deviation from \cite{Carlson:1992fn} in the numerical constants comes from a slightly improved value of $N_{\text{eff}}$)
\begin{equation}
  \frac{\Lambda}{T_d'} + 2\, \ln\left(\frac{\Lambda}{T_d'}\right) = \frac{3}{4} \ln 
\left( \frac{\lambda g'^{7/4}}{\Omega h^2}\right)- \frac{5}{4} \ln(g'^{1/4}\xi)+43.4\,.\label{eq:Tdrad}
\end{equation}
In the case where the universe is matter dominated at decoupling,
$T_d'$ is determined by
\begin{equation}
  \frac{\Lambda}{T_d'} + \frac{3}{2}\, \ln\left(\frac{\Lambda}{T_d'}\right) = \frac{2}{3} \ln 
\left( \frac{\lambda g'^{7/4}}{\Omega h^2}\right)- \frac{2}{3} \ln(g'^{1/4}\xi)+38.07\,.
\label{eq:Tdmat}
\end{equation}
The appearance of $\Lambda$ is a substitute for the dark matter mass
$m'$ of \cite{Carlson:1992fn}.  This is motivated by the fact that
glueballs are expected to have mass $m'=c \Lambda$ with $c\gtrsim 1$
an $O(1)$ coefficient. We take $c=1$ for simplicity, since it does not
significantly affect our conclusions.

Using these results, \cite{Carlson:1992fn} studied the implications of
the observed dark matter relic abundance from decoupling.

\vspace{.5cm} We instead take an ultraviolet perspective, where high
scale physics such as moduli stabilization in string theory could set
a wide range of values for the ultraviolet gauge coupling
$\alpha_{\text{UV}}$; $\Lambda$ depends exponentially on $\alpha_{\text{UV}}$. We
will see that the glueball relic abundance is linear in $\Lambda$ to a
good approximation, and therefore the glueball is a poor dark matter
candidate since $\alpha_{\text{UV}}$ must be exponentially fine tuned to obtain a relic
abundance close to the observed value. However, we will see that dark
glueballs can place strong constraints on the ultraviolet theory.

Let us compute the relic abundance in terms of the confinement scale
rather than the decoupling temperature. To do so, we use
(\ref{eq:relicvsTd}) to trade $T_d'$ for the relic abundance in
(\ref{eq:Tdrad})-(\ref{eq:Tdmat}). In the case of radiation domination
at decoupling this leads to
\begin{equation}
\Omega h^2 \!=\! \frac{\Lambda}{3.6\, {\rm eV}} \frac{4}{5\, \xi \,\,W\!\left(7.45\!\times\! 10^{12} f^{6/5} g^{4/5} \xi^{-2/5}\left(\frac{3.6\, {\rm eV}}{\Lambda}\right)^{3/5}\right)}
\label{eqn:relicrad}
\end{equation}
where $W(x)$ is the Lambert $W$-function or product logarithm, which is the inverse of $f(x)=xe^x$
much as log is the inverse of $f(x)=e^x$. In the case of matter domination at 
decoupling the relic abundance is
\begin{equation}
\Omega h^2 = \frac{\Lambda}{3.6\, {\rm eV}} \frac{6}{5\, \xi \,\,W\left(1.28\times 10^{17} f^{8/5} g^{6/5} \left(\frac{3.6\, {\rm eV}}{\Lambda}\right)^{4/5}\right)}
\label{eqn:relicmat}
\end{equation}
These calculations are  valid for $\Lambda / T_d' > 1$.

What is the relic abundance outside of this regime?  Naively
considering $\Lambda / T_d' < 1$ is not physically sensible, since for
temperatures $T' > \Lambda$ the dark sector is comprised of
relativistic gluons and the effective field theory in which $3\to 2$
interactions were computed is not valid. Instead, as the universe
cools in this other regime glueballs form and immediately decouple,
i.e.\  $T_d'\simeq \Lambda$. This together with (\ref{eq:relicvsTd})
gives a relic abundance
\begin{equation}
\Omega h^2 \simeq \frac{\Lambda }{3.6\, {\rm eV}\, \xi}\,,
\label{eqn:relicno3to2}
\end{equation}
which closely matches the results of \cite{Boddy:2014yra}, which set
$3\to 2$ interactions to zero. 

\vspace{.2cm} How strongly do $3\to 2$ interactions affect the relic
abundances (\ref{eqn:relicrad})-(\ref{eqn:relicmat})? Specifically,
how much do they deplete the relic abundance (\ref{eqn:relicno3to2})
that would be obtained in the absence of these interactions?  This can
be approximated by noting the mild dependence of $W(x)$ on $x\in
\mathbb{R}_{>1}$, which is similar to the mild dependence of $\log(x)$
on similar $x$. For example, though $W(10)$ is $\cO(1)$,
$W(10^{1000})$ is only $\cO(10^3)$. Since both (\ref{eqn:relicrad})
and (\ref{eqn:relicmat}) have $\Omega h^2 \simeq (\Lambda / {\rm eV})
\times (1/ \xi \, W(x))$, the order of magnitude of the dark glueball
relic abundance is primarily set by $\Lambda$ and $\xi$.

In particular, if (\ref{eqn:relicno3to2}) oversaturates the observed
relic abundance by many orders of magnitude, $3\to 2$ interactions
cannot ameliorate the situation.

\vspace{.2cm}
\noindent {\emph{Overproduction of Democratic Dark Glueballs}}
 
\begin{figure*}[htb]
  \includegraphics[width=.95\columnwidth]{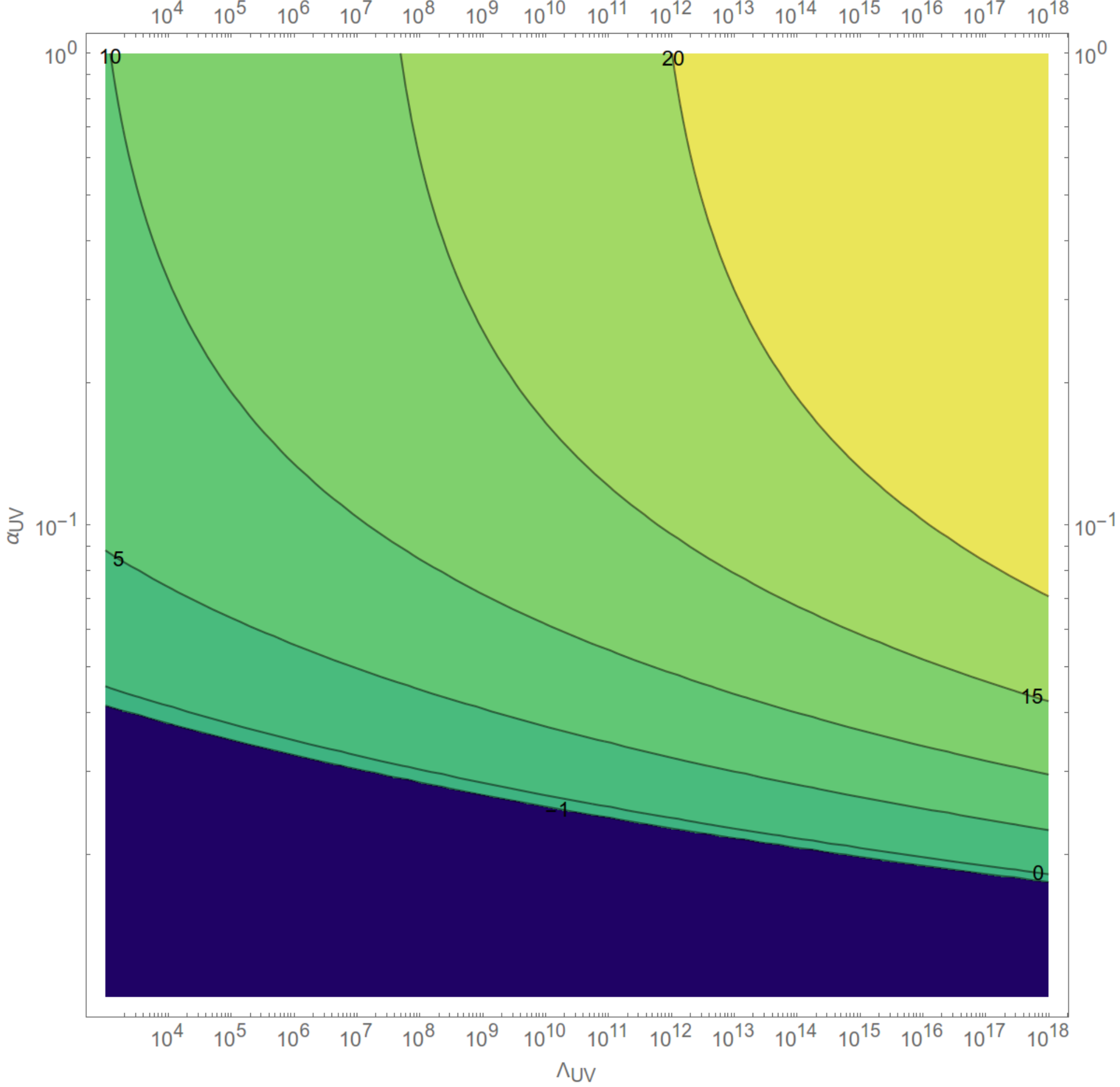}
  \includegraphics[width=.95\columnwidth]{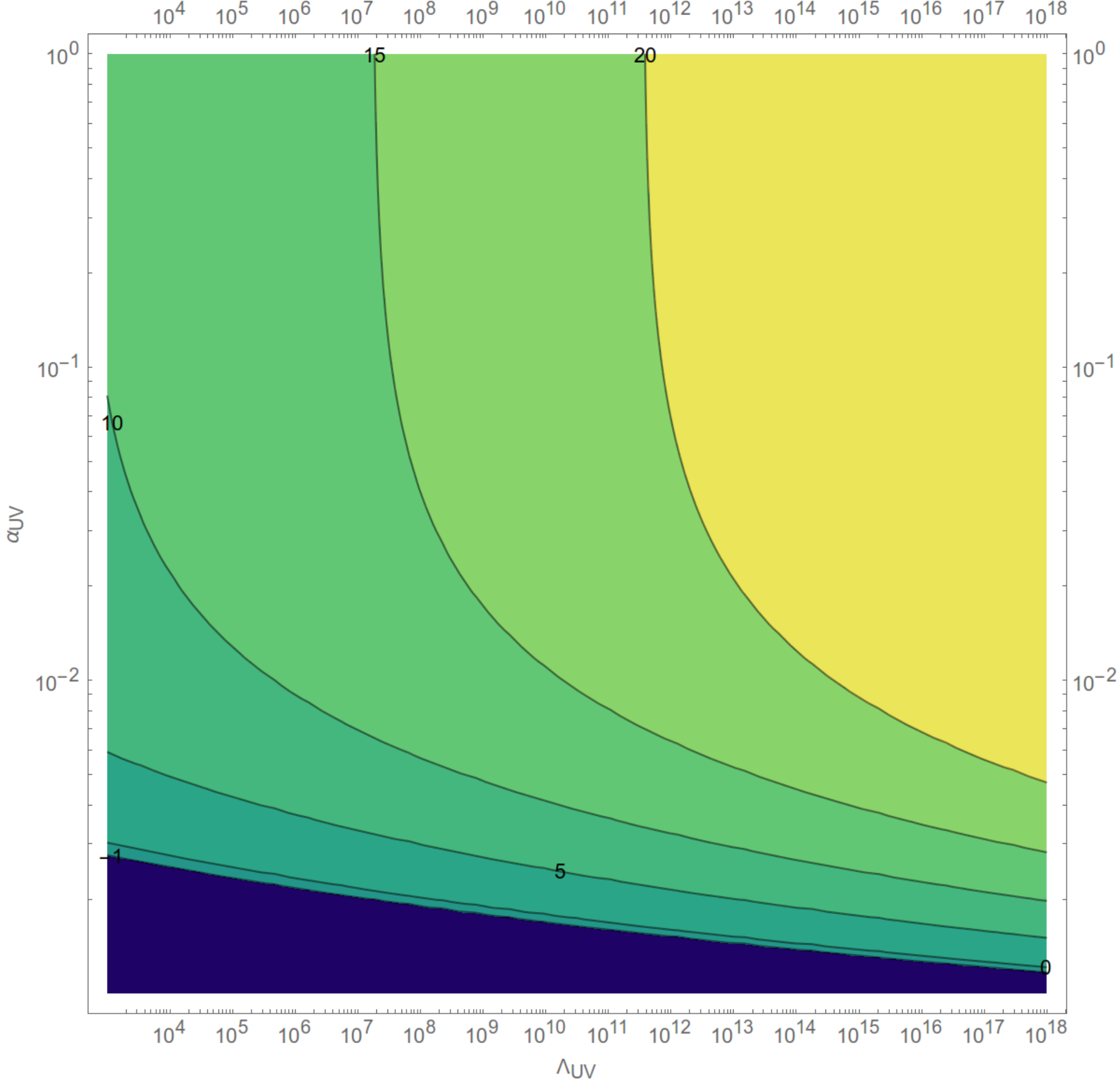} \\
  \includegraphics[width=.95\columnwidth]{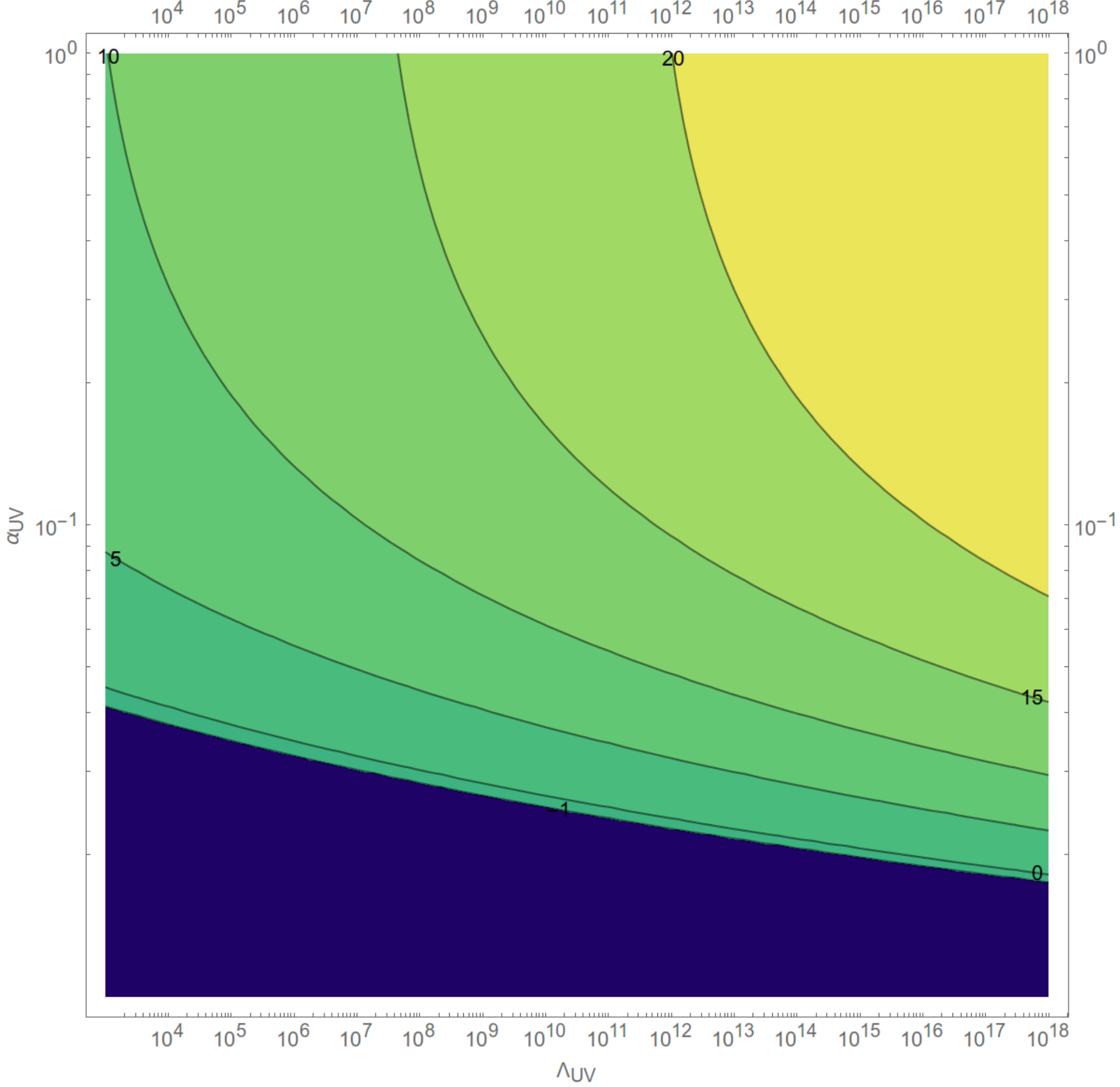}
  \includegraphics[width=.95\columnwidth]{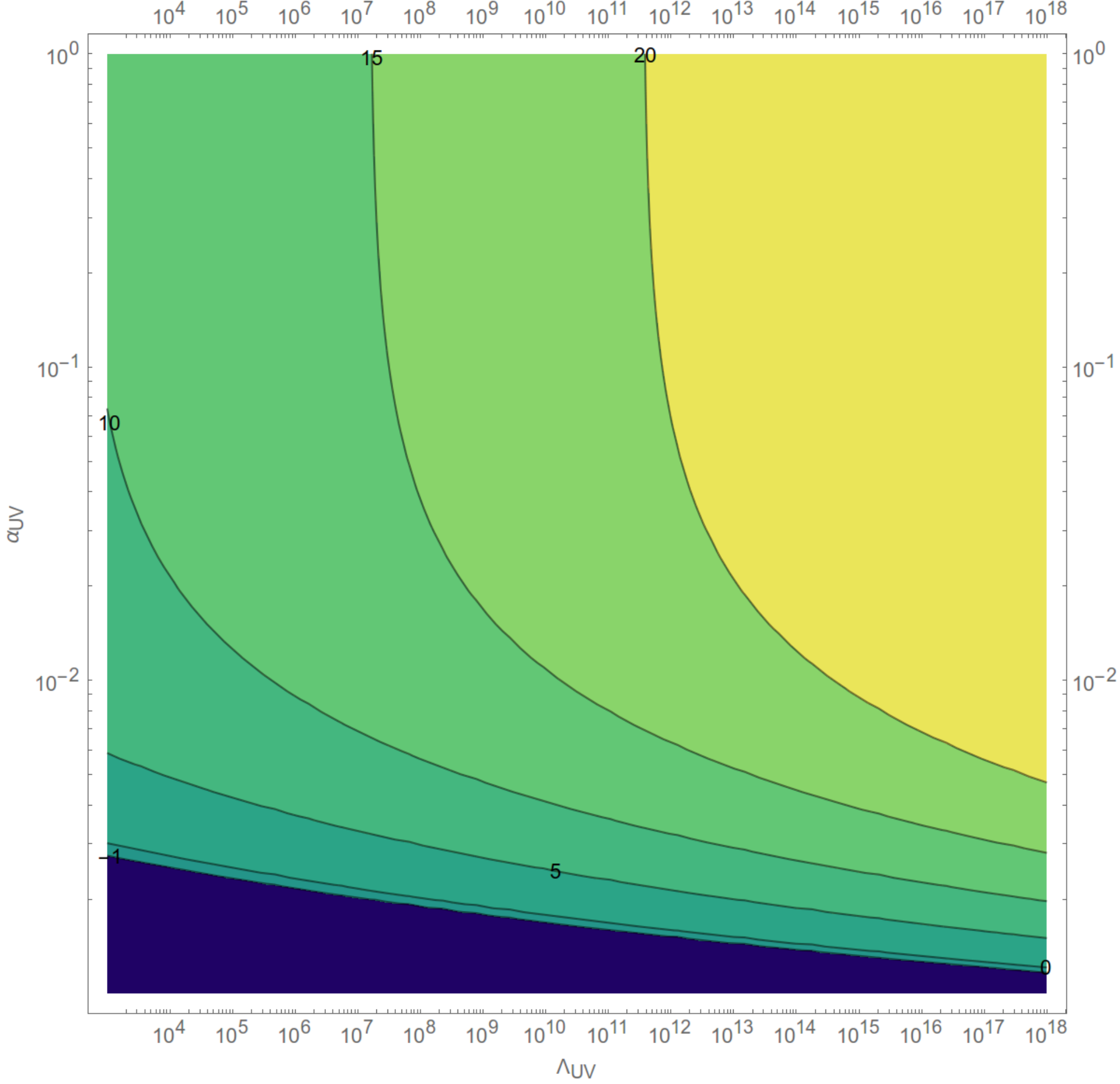}
  \caption{Glueball relic abundance as a function of $\alpha_{\text{UV}}$ and $\Lambda_{\text{UV}}$ for with $\xi=1$, $f=.1$.
   In each figure the relic abundance is oversaturated outside of the blue region.
   \emph{Upper left:} $G=SU(2)$ and radiation domination at decoupling.
   \emph{Upper right:} $G=E_8$ and radiation domination at decoupling.
   \emph{Lower left:} $G=SU(2)$ and matter domination at decoupling.
   \emph{Lower right:} $G=E_8$ and matter domination at decoupling.
  }
  \label{fig:glueAlphaVsLambdaRelics}
\end{figure*}

Let us study the glueball relic abundance in the democratic case
$\xi=1$, focusing on its dependence on $\Lambda$ and $\xi$ for natural
values of ultraviolet parameters.

We compute $\Lambda$ via the beta function of super Yang-Mills theory,
which gives
\begin{equation}
\Lambda \equiv \Lambda_{\text{IR}} =  \Lambda_{\text{UV}}e^{-\frac{2\pi}{3C_2(G)\,\alpha_{\text{UV}}}}\,,
\end{equation}
where $C_2(G)$ is the dual Coxeter number of the gauge group $G$,
$\alpha_{\text{UV}}$ is the ultraviolet gauge coupling evaluated at scale
$\Lambda_{\text{UV}}$, and $\Lambda_{\text{IR}}$ is the scale at which $\alpha$
diverges. We use the supersymmetric beta functions all the way down to
low scale for both simplicity and generosity. The former applies since
this choice avoids the introduction of the scale $\Lambda_{\text{SUSY}}$, and
the latter applies since the supersymmetric beta functions give rise
to lower confinement scales; the oversaturation problem that we will
encounter is only exacerbated by using non-supersymmetric beta
functions below $\Lambda_{\text{SUSY}}$.

The groups that we study are $SU(2)$, $SU(3)$, $G_2$, $SO(7)$,
$SU(5)$, $SO(8)$, $SO(10)$, $F_4$, $E_6$, $E_7$, and $E_8$, which are
two of the most commonly studied grand unified groups ($SU(5)$ and
$SO(10)$) together with the group factors that may appear
geometrically for general values of complex structure moduli in $d=4$
F-theory \cite{Grassi:2014zxa}.  These groups have $C_2(G)$ given by
$2$, $3$, $4$, $5$, $5$, $6$, $8$, $9$, $12$, $18$, and $30$,
respectively. These values imply that, for fixed $\Lambda_{\text{UV}}$, a
change from one group to another can give rise to the same $\Lambda$
by an $O(1)$-$O(10)$ change in $\alpha_{\text{UV}}$. The same relic abundance
for glueballs of different group can therefore be obtained by a
relatively small $\alpha_{\text{UV}}$ change.

In Figure \ref{fig:glueAlphaVsLambdaRelics} we take $G=SU(2)$ and
$E_8$ glueballs as prototypes, since they have the lowest and highest
confinement scales, respectively, for fixed $\Lambda_{\text{UV}}$ and
$\alpha_{\text{UV}}$. The relic abundances are computed in both the case of
radiation domination and matter domination at decoupling, taking
$\xi=1$ and studying the natural parameter space $10^{-3}\leq
\alpha_{\text{UV}} \leq 1$, $10^{3}\, {\rm GeV} \leq \Lambda_{\text{UV}} \leq
10^{18}\, {\rm GeV}$.  On a log-log scale we see that there is little
difference between the two cases. The blue band represents
undersaturation of the observed relic abundance \cite{Ade:2015xua}
$\Omega_{\text{obs}} h^2 = 0.1199 \pm 0.0027$, with saturation occurring at
the edge.  The $\Omega h^2 = 1$ contour sits very close to the
observed relic abundance contour, and the $\Omega h^2 = 10^5, 10^{10},
10^{15}, 10^{20}$ contours make up the remaining parameter space. This
figure demonstrates that for $\xi=1$, the smallest ($SU(2)$) and
largest ($E_8$) glueball relic abundances for these groups
oversaturate the observed value by many orders of magnitude for much
of this parameter space.

\begin{figure*}[th]
  \includegraphics[width=.8\columnwidth]{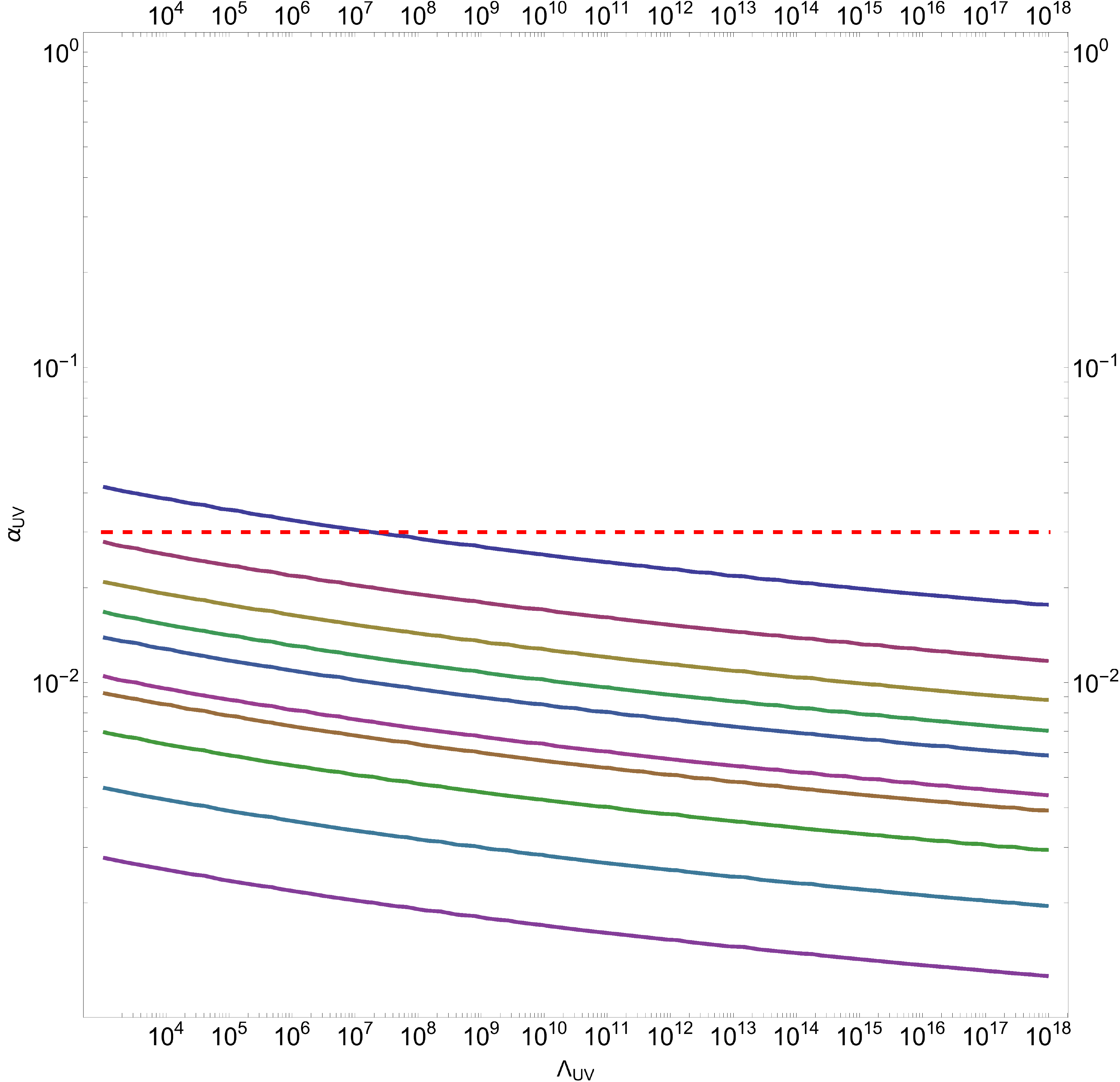} 
  \includegraphics[width=.8\columnwidth]{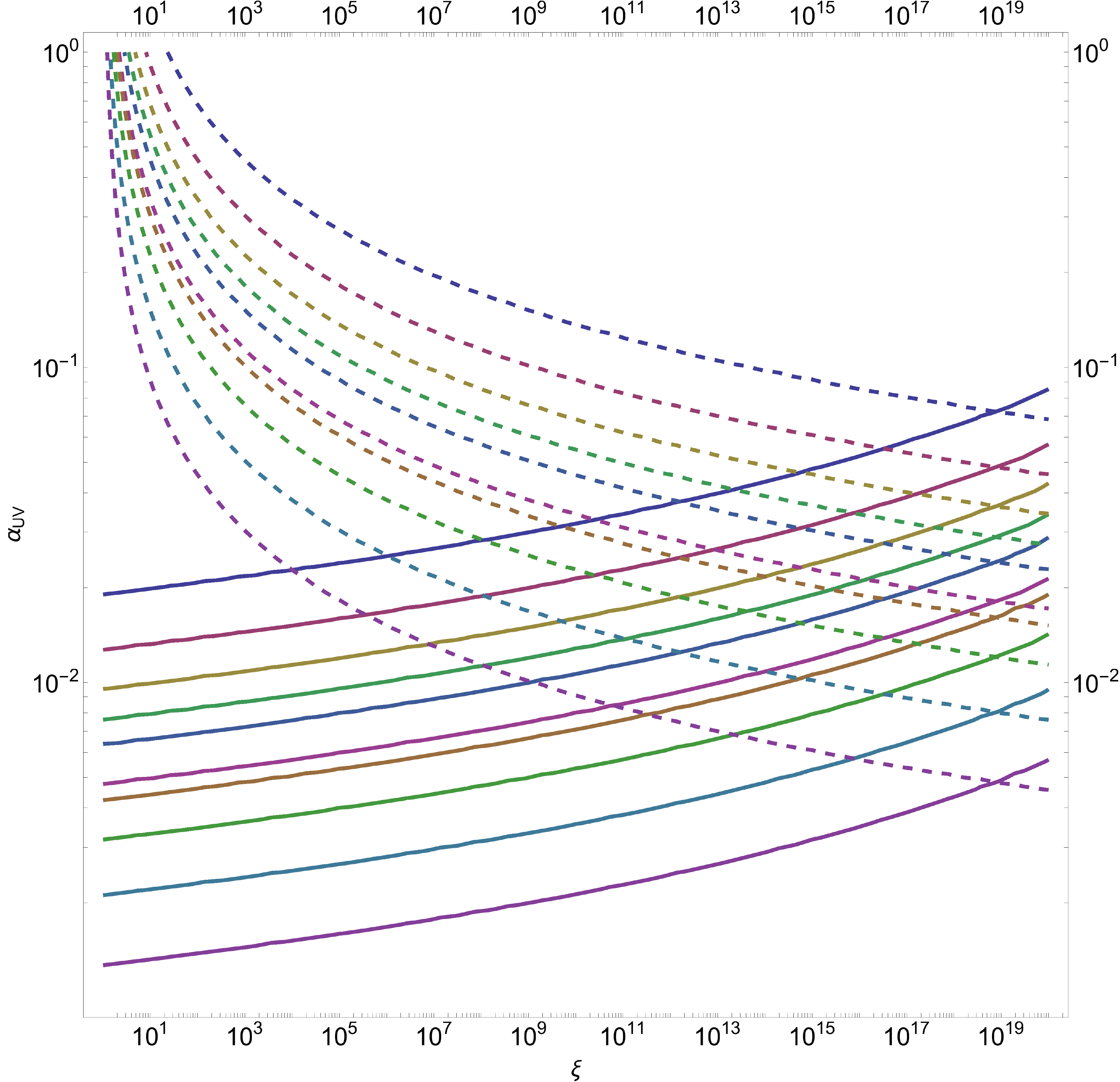} 
  \hspace{.5cm}
  \includegraphics[width=.27\columnwidth]{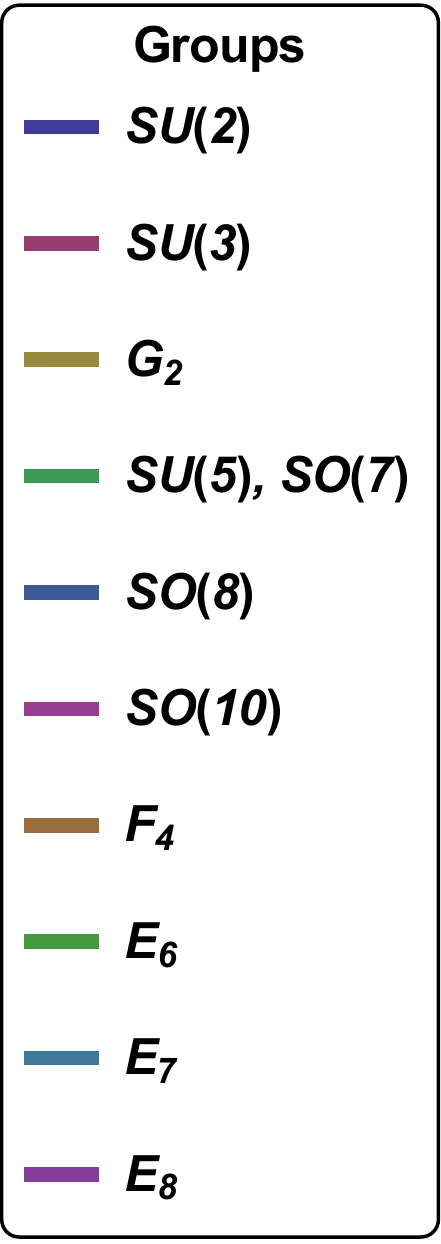} 
  \caption{{\em Left:} Contours where the glueball relic abundance
    saturates the observed relic abundance for various groups as a
    function of $\alpha_{\text{UV}}$ and $\Lambda_{\text{UV}}$ (in GeV), with oversaturation
    occurring in the region above each contour.  The horizontal red
    line marks $\alpha_{\text{UV}}=\alpha_{\text{GUT}}$. {\em Right:} The glueball
    relic abundance matches the observed relic abundance along dark
    contours, with oversaturation above. To the right of the dashed
    contours the dark sector is not reheated above its confinement
    scale.}
  \label{fig:glueRelicObservedManyGroups}
\end{figure*}

Having demonstrated how rapidly $\Omega h^2$ increases in the
$\Lambda_{\text{UV}}$-$\alpha_{\text{UV}}$ plane, in Figure
\ref{fig:glueRelicObservedManyGroups} we present the contours on which
each of the groups we study saturates the observed relic abundance. Over half
of the parameter space is ruled out for all of the groups, and for
some groups it is much more. For $\Lambda_{\text{UV}} > 10^{9}\, {\rm GeV}$
the observed relic abundance is oversaturated for
$\alpha=\alpha_{\text{GUT}}\simeq .03$ for all groups,
though moduli stabilization may fix the ultraviolet gauge coupling at
significantly different values. 

In conclusion, stable dark glueballs in the
democratic scenario $\xi=1$ oversaturate the relic abundance for much
of the ultraviolet parameter space, putting strong constraints on
ultraviolet theories that realize dark Yang-Mills sectors. Henceforth
we will call this the dark glueball problem, for brevity, and in the
next two sections we will study mechanisms that could potentially
solve it.

\vspace{.5cm}\noindent\textbf{Constraints from Preferential Reheating.}

One potential solution to the dark glueball problem is to reheat
preferentially into the visible sector, constraining models of
reheating via inflaton or modulus decay.  Preferential reheating into
the visible sector, i.e.\ $\xi > 1$, may solve the problem by either
depleting the dark glueball relic abundance or by leaving the regime
of validity $T_{\text{rh}}' \gtrsim \Lambda$ for the production mechanism of
\cite{Carlson:1992fn}.

Let us study the former by deriving bounds on $\xi$ that are
sufficient to not oversaturate the observed dark matter relic
abundance, beginning with models that do not exhibit $3\to 2$
interactions. This will give an approximate lower bound on $\xi$ that
is rough, but good enough for some purposes since $3\to 2$
interactions cannot significantly suppress a very large relic
abundance. The bound that avoids oversaturation is
\begin{equation}
\xi \gtrsim \frac{\Lambda}{3.6\, {\rm eV}\, \Omega_{\text{obs}}h^2} = 2.3 \frac{\Lambda}{{\rm eV}}\,.
\label{eqn:boundno3to2}
\end{equation}
Thus, in the absence of $3\to 2$ interactions, confinement scales
$\Lambda \gtrsim 1\, {\rm eV}$ require there to be more entropy in the
visible sector, i.e.\ $\xi > 1$. As a reference point, hidden sectors
with confinement scales $\Lambda = \Lambda_{\text{QCD}} \simeq 10^6\, {\rm
  eV}$ require $\xi \gtrsim 10^6$.  The bounds are weakest for lower
rank groups, since they have lower confinement scales, but the
constraint can be significant even for low rank groups. For example,
with $\alpha=\alpha_{\text{GUT}}\simeq .03$ and $G=SU(2)$,
 the approximate bound is $\xi \gtrsim 1.5\times
10^{10}$. We will study the accuracy of this approximate bound
momentarily by taking into account $3\to 2$ interactions.

Alternatively, the dark glueball problem may be solved if the
production mechanism of \cite{Carlson:1992fn} is not in effect. This
arises as follows. The visible sector reheat temperature after
inflation (or modulus decay) must satisfy $T_{\text{rh}} \lesssim M_{\text{GUT}}$.
The relationship $\xi = s/s' = g_S T^3 / (g_S' T'^3)$ implies $T_{rh}'
= (g_{S,\text{rh}}/(g_{S,\text{rh}}'\xi))^{1/3} T_{\text{rh}}\simeq \xi^{-1/3} T_{\text{rh}}$,
where the latter approximation gives a gauge group independent
relationship that will suffice for our purposes since the
$g_S$-dependence will make little qualitative difference on a log-log
scale. Then the bounds $T_{\text{rh}}' \gtrsim \Lambda$ and $T_{\text{rh}} \lesssim
M_{\text{GUT}}$ together imply $\xi \lesssim (M_{\text{GUT}}/\Lambda)^3$, so that
the bound associated with leaving the regime of validity for glueball
production is
\begin{equation}
\xi \gtrsim \left(\frac{M_{\text{GUT}}}{\Lambda}\right)^3\,.
\label{eqn:RHbound}
\end{equation}
For confinement scales that we study $\Lambda < M_{\text{GUT}}$ and this
bound implies that glueball production is valid for $\xi \leq 1$,
which includes the democratic scenario. If $\xi$ is increased from
$\xi=1$ with fixed $\Lambda$, however, eventually the bound will be
satisfied, in which case the dark sector reheats to a temperature
below the confinement scale and the glueball production mechanism we
study is not in effect. Other production mechanisms may potentially
arise, most plausibly when $T_{\text{rh}}'\simeq \Lambda$, but we will leave
such studies to future work and will clearly delineate regions of
parameters space where (\ref{eqn:RHbound}) is violated.

Summarizing, if either of the bounds (\ref{eqn:boundno3to2}) or
(\ref{eqn:RHbound}) are satisfied then the glueball relic abundance is
not oversaturated.

\vspace{.5cm}

Let us see when these bounds are satisfied for various groups and
values of $\alpha_{\text{UV}}$, fixing $\Lambda_{\text{UV}}=10^{16}\, {\rm GeV}$.
The bound (\ref{eqn:boundno3to2}) can be slightly weakened by the
incorporation of $3\to 2$ interactions, which are taken into account
in the right panel of Figure~\ref{fig:glueRelicObservedManyGroups} with $f=0.1$. The
solid and dashed lines are those in which the bounds for the relic
abundance (with $3\to 2$ interactions) and regime of validity are
saturated, respectively, for a particular group. Oversaturation occurs
for a glueball with fixed $G$ for points in the parameter space above
the associated solid contour, but below the associated dashed contour. 

Both bounds must be taken into account:
for example, for fixed $\alpha \gtrsim .1$ satisfying the relic
abundance bound would require $\xi \gtrsim 10^{20}$, but the regime of
validity bound may be satisfied for smaller values of $\xi$, avoiding
oversaturation. Conversely, for fixed $\alpha_{\text{UV}}\lesssim
7\times 10^{-3}$ there are values of $\xi$ that violate the relic
abundance bound but not the regime of validity bound. For any fixed
$\alpha_{\text{UV}}$ and $G$ the minimum value of $\xi$ sufficient to avoid
the dark glueball problem can be read from the associated solid and
dashed contours. For fixed $G$, the value of $\alpha_{\text{UV}}$ that
requires the largest $\xi$ to satisfy the bounds occurs when both
bounds are saturated, which occurs for the $\alpha_{\text{UV}}$ at which the
associated solid and dashed contours intersect. Interestingly, this
always occurs for $.005 < \alpha_{\text{UV}} < .1$, which is a range that
contains $\alpha_{\text{GUT}}$.

\vspace{.5cm}\noindent\textbf{Constraints from Decays to Moduli and Axions.}

An additional mechanism for evading the consequences of the above
analysis is to allow the glueballs to decay to lighter degrees of
freedom. In the present context we are assuming a hidden sector devoid
of matter charged under the confining group, and we do not assume a
coupling (or `portal') to the fields of the Standard Model~\cite{Acharya:2016fge}. This
leaves only potentially light moduli and/or axionic fields as decay
channels.

We have in mind the geometrical moduli generic to all string
compactifications.  While their masses are {\em a priori}
undetermined, general arguments in supergravity~\cite{Denef:2004cf,GomezReino:2006dk,Acharya:2010af} 
suggest that the lightest such modulus ought to have a mass
comparable to that of the gravitino mass -- roughly the size of the
soft scalar masses in the observable sector -- and thus (presumably)
on the order of 10\,TeV. Argument from the successful predictions of Big
Bang Nucleosynthesis (BBN) imply that the masses of these moduli must
be no less than approximately 50\,TeV, and we will take this number as
a benchmark throughout the remainder of the paper. Note that the
argument from BBN persists even in the absence of low-energy
supersymmetry.

Let us again denote the glueball in the low energy effective field
theory by $\phi$ and designate its mass by $m_{\phi}$, where
$m_\phi \simeq \Lambda$. Let us denote a generic modulus field as
$\chi$. Then assuming the decay into such moduli ($\phi \to \chi\chi$)
is kinematically accessible, we can estimate the lifetime by utilizing
a dimension-six operator such as
\begin{equation}
\mathcal{O}_{6} = \frac{{\rm Tr}(G^a_{\mu\nu}G_a^{\mu\nu}) \chi\chi}{M_s^2} \to \frac{\Lambda^3}{M_s^2} \phi \chi\chi \, ,
\label{eq:dim6}
\end{equation}
where the trace is over the gauge degrees of freedom, and we replace
the field strengths with $\phi \Lambda^3$ in the effective theory
below the confinement scale. The scale $M_s$ is the scale at which the
supergravity effective theory is valid. We will take $M_s = M_{\rm
  GUT}$ in explicit computations.

The width associated with~(\ref{eq:dim6}) is given by
\begin{equation}
\Gamma_{6} = \frac{1}{4\pi}\frac{1}{m_{\phi}}\left(\frac{\Lambda^3}{M_s^2}\right)^2 \sqrt{1-\frac{4m_{\chi}^2}{m_{\phi}^2}}\, ,
\label{eq:Gam6}
\end{equation}
where $m_{\chi}$ is the modulus mass. Successful BBN requires that the universe be radiation dominated at the time that the relative abundances of protons and neutrons are set, roughly $0.1\,{\rm s}$. As the glueballs quickly come to dominate the energy density of the universe upon confinement, we must therefore demand that the lifetime associated with~(\ref{eq:Gam6}) be no longer than this value. Taking $m_{\chi} = 50\,{\rm TeV}$, $M_s = M_{\rm GUT} = 10^{16}\,{\rm GeV}$ and $m_{\phi} = \Lambda$, this implies a constraint
\begin{equation} \Lambda \geq 2.4 \times 10^8\, {\rm GeV} \,, 
\label{eq:Lambdabound}
\end{equation}
Glueballs with masses below the bound in~(\ref{eq:Lambdabound}) but
above $2m_\chi\simeq 100\, {\rm TeV}$ decay after the onset of BBN and
spoil its successful predictions.

Alternatively one might hope that
decays into even lighter objects could remedy the situation. A
well-motivated candidate would be a light axionic field. Such states
are common in string theory; indeed, the moduli fields themselves will
have an imaginary component that behaves like an axion in the
low-energy effective theory.

For our purposes it is sufficient to consider a single such axionic
state, with an associated decay constant $f_a$. We will assume all
interactions between the glueball and the axion are suppressed by this
scale. In practice, one commonly finds $f_a \simeq M_s$ in typical
string models, but we will be agnostic as to the precise value of this
constant.  Prior to taking into account non-perturbative effects, the
axion enjoys a shift symmetry and therefore only appears in the
Lagrangian through derivative interactions. Thus an operator such
as~(\ref{eq:dim6}) is forbidden, and one must instead turn to a
dimension-eight interaction governed by
\begin{equation} 
\mathcal{O}_{8} = \frac{{\rm Tr}(G^a_{\mu\nu}G_a^{\mu\nu}) \partial_{\rho} a\, \partial^{\rho} a}{f_a^4} \to \frac{\Lambda^3}{f_a^4} \phi\, \partial_{\rho} a\, \partial^{\rho} a \, ,
\label{eq:dim8}
\end{equation}
resulting in a decay width given by
\begin{equation}
\Gamma_{8} = \frac{1}{64\pi}m_{\phi}^3\left(\frac{\Lambda^3}{f_a^4}\right)^2 \simeq \frac{\Lambda}{64\pi}\left(\frac{\Lambda}{f_a}\right)^8 \, ,
\label{eq:Gam8}
\end{equation}
where we are making the assumption $m_a \ll m_{\phi}$, for
simplicity. The eight powers of $\Lambda/f_a$ suppression makes the
resulting glueball lifetime fantastically long. For example, a
glueball whose confinement scale saturated the bound
in~(\ref{eq:Lambdabound}) would require
\begin{equation} f_a \leq 1.1 \times 10^{12}\, {\rm GeV}
\label{eq:fabound}
\end{equation}
to allow for a decay before the onset of BBN. A glueball with
confinement scale $\Lambda = 1\,{\rm TeV}$ would require $f_a \leq 9.7
\times 10^{5}\,{\rm GeV}$ to decay before BBN. Fixing $f_a = M_s =
10^{16}\,{\rm GeV}$, we find $\tau(\phi \to a a) > 0.1\,{\rm s}$ for
all confinement scales below $8\times 10^{11}\, {\rm GeV}$, showing that
those cases which cannot decay to geometrical moduli promptly enough
will not be rescued by decays to axions if $f_a\simeq 10^{16}\, {\rm
  GeV}$.

Under these assumptions we have established that hidden Yang-Mills
sectors with $\Lambda \gtrsim 3\, {\rm eV}$ are cosmologically ruled
out unless the upper bound on $f_a$ is satisfied for $50\,{\rm
  TeV}\lesssim \Lambda \lesssim 10^9\, {\rm GeV}$, the confinement
scale is quite high ($\Lambda \gtrsim10^9\,{\rm GeV}$), or an extreme
measure of preferential reheating is engineered. 

\vspace{.3cm}
One might argue the
latter constraint could be relaxed if faster decay rates could be
motivated. We have chosen the operators~(\ref{eq:dim6})
and~(\ref{eq:dim8}) to reflect the underlying structure of the
UV~theory above the scale of confinement. However, treating the
glueball as a gauge-singlet scalar does naively allow the operator
\begin{equation} \mathcal{O}_5 = \frac{\phi}{M_s} \partial_{\mu} \chi \partial^{\mu} \chi \to \Gamma_5 =   \frac{1}{64\pi}\frac{m_{\phi}^3}{M_s^2}\sqrt{1-\frac{4m_{\chi}^2}{m_{\phi}^2}}  \, ,
\label{eq:dim5}
\end{equation}
which could mediate glueball decay to moduli fields. The
operator in~(\ref{eq:dim5}) allows for prompt ($\tau < 0.1\,{\rm s}$)
decays of glueballs to moduli for all cases in which the decay is
kinematically accessible, assuming a modulus mass $m_{\chi} = 50\,{\rm
  TeV}$ and $M_s = 10^{16}\,{\rm GeV}$. This is to be expected: the
operator in~(\ref{eq:dim5}) is precisely the operator that is
postulated to allow sufficiently heavy moduli to decay into light
degrees of freedom prior to BBN, thereby solving the cosmological
moduli problem. \footnote{One might choose to suppress the operator
  in~(\ref{eq:dim5}) by the confinement scale $\Lambda$, as opposed to
  the much larger scale $M_s$. Or one might argue that a
  superrenormalizable coupling like $\Lambda \phi \chi\chi$ be
  utilized. Both modifications would only serve to shorten the
  lifetime, so we do not consider them here.}  
For much lighter
glueball states (lower confinement scales), the moduli decay channel
closes and~(\ref{eq:dim5}) with $\chi \to a$ and $M_s\to f_a$ now governs decay to axions (where we might
assume $m_a \ll m_{\phi}$). In order to keep the lifetime sufficiently
short, we must now lower the value of $f_a$ away from $10^{16}$ to
compensate for the much lower glueball mass. For example, taking an
intermediate value $f_a = 10^{12}\,{\rm GeV}$, we find that the
glueball decays sufficiently quickly into massless axions provided
$\Lambda > 11\,{\rm GeV}$. A glueball whose mass is comparable to the
QCD scale ($\Lambda = 100\,{\rm MeV}$) would require $f_a < 9\times
10^{8} \, {\rm GeV}$ to decay into axions during the brief window
between confinement and the beginning of BBN.

However, we do not believe that
(\ref{eq:dim5}) is present due to the structure of dark
gluon interactions with moduli in the UV theory and the emergent nature of
the glueball $\phi$ below the confinement scale.

\begin{figure}[th]
  \centering
  \includegraphics[width=.9\columnwidth]{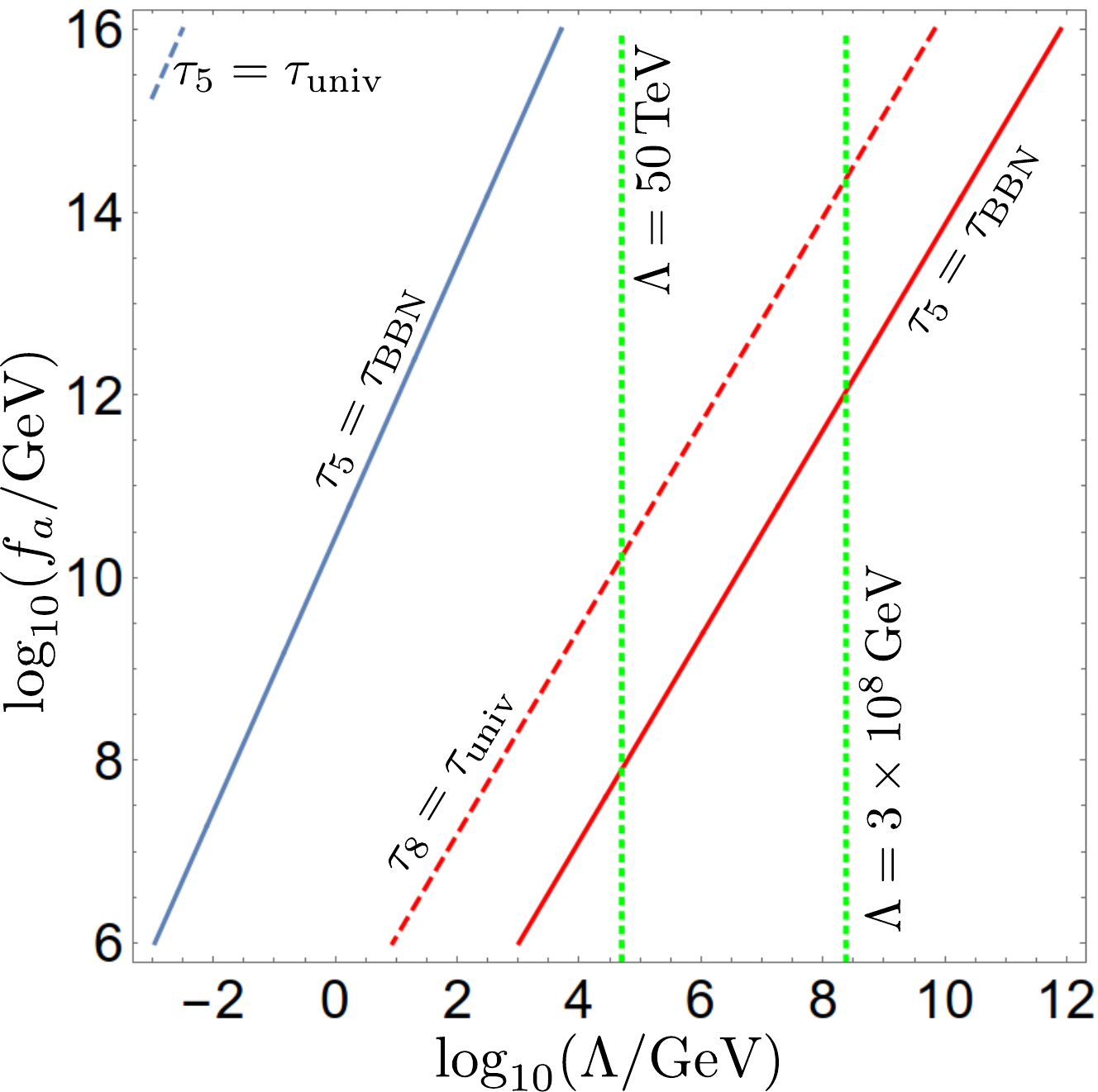} 
   \caption{Glueball lifetimes as a function of $f_a$ and $\Lambda$, with contours for $\tau_n=\tau_{\rm univ}$
and $\tau_n=\tau_{\rm BBN}=0.1s$ for $n=5,8$ corresponding
to decay via dimension $5$ and $8$ operators.}
   \label{fig:Lifetimes}
\end{figure}

Results on axion decays are summarized in Figure \ref{fig:Lifetimes}.  In the case of decay via
the dimension eight (five) operator the glueball is stable and
oversaturates the observed relic abundance to the left of the red (blue) dashed
line. It is unstable but spoils nucleosynthesis between the red (blue)
dashed and solid lines. To the right of the red (blue) solid line the
glueball is unstable and decays prior to nucleosynthesis.

\vspace{.5cm}\noindent\textbf{Constraints from Dark Radiation.}

In the better motivated case in which decay operators arise from the underlying structure of the hidden gauge fields, we find that decays to moduli require a confinement scale above $10^8\, {\rm GeV}$ and prompt decays to axions are possible for lower confinement scales if the axion
decay constant is sufficiently small.
If the decay operators descend from `naive' effective field theory, the glueball will decay into (cosmologically safe) moduli, prior to BBN, provided that such decays are kinematically accessible ({\em i.e.}\ $m_{\phi} \simeq 2 m_{\chi} \simeq 100\,{\rm TeV}$). Furthermore, decays into light axions are possible for confinement scales between the modulus mass and the onset of BBN, provided the axion decay constant takes intermediate values $10^{10}\,{\rm GeV} \lappeq f_a \lappeq 10^{16}\,{\rm GeV}$.  However, these axionic decay products are not necessarily cosmologically benign.

The prediction from BBN for the abundance of $^4$He is sensitive to the Hubble parameter at the time BBN begins, and is thus sensitive to the number of relativistic degrees of freedom present in the cosmos at that time,  $g_*^{\rm hid}|_{\rm BBN}$~\cite{Fields:2006ga}. The value of $g_{*}^{\rm hid}|_{\rm CMB}$ is constrained again at the time of CMB formation~\cite{Ade:2015xua}. The upper bounds, at the 95\% confidence, on these quantities are given by~\cite{Kane:2015qea}
\begin{eqnarray} g_{*}^{\rm hid}\big|_{\rm BBN}   & \leq & 2.52\, \xi_T^{4}\big|_{\rm BBN}\,,  \nonumber \\ 
 g_{*}^{\rm hid}\big|_{\rm CMB} & \leq & 0.18\, \xi_T^{4}\big|_{\rm CMB}\, ,
\label{eq:gstarbound}
\end{eqnarray}
where we recall that $\xi_T = T/T'$.

Any non-Abelian group which avoids overproduction of glueball dark matter by achieving a very low confinement scale, governed by~(\ref{eqn:relicno3to2}), would therefore have a potential problem from dark radiation at the time of BBN ($T_{\rm BBN} \simeq 1\,{\rm MeV}$) and/or the time at CMB formation ($T_{\rm CMB} \simeq 0.25\,{\rm eV}$). The situation is ameliorated somewhat by the fact that the visible sector has fewer degrees of freedom at those late times than it did at some earlier, primordial time~\cite{Feng:2008mu}. Thus the visible sector should be `warmer' than the hidden sector at the time of BBN and CMB formation (assuming the hidden sector has remained a plasma of relativistic gluons throughout this history). Indeed one expects
\begin{eqnarray}
\xi_T^3\big|_{\rm BBN} &=& \xi_T^3 \big|_{\rm rh} \left(\frac{g_{*s}^{\rm vis} (T_{\rm rh})}{g_{*s}^{\rm vis} (T_{\rm BBN})}\right) \nonumber\\
		       &=& \xi_T^3 \big|_{\rm rh} \left(\frac{g_{*s}^{\rm vis} (T_{\rm rh})}{10.75}\right)\,,  \label{eq:xiBBN} \\
\xi_T^3\big|_{\rm CMB} &=& \xi_T^3 \big|_{\rm rh} \left(\frac{g_{*s}^{\rm vis} (T_{\rm rh})}{g_{*s}^{\rm vis} (T_{\rm CMB})}\right) \nonumber\\
		       &=& \xi_T^3 \big|_{\rm rh} \left(\frac{g_{*s}^{\rm vis} (T_{\rm rh})}{3.36}\right)\, ,  \label{eq:xiCMB}
\end{eqnarray}
where $\xi_T|_{\rm rh}$ is the ratio of temperatures at some
primordial scale at which both the visible and hidden sectors are
populated through some sort of reheating. Assuming the reheat
temperature is such as to populate the entire MSSM~field content, one
finds from~(\ref{eq:gstarbound})
\begin{eqnarray} g_{*}^{\rm hid}\big|_{\rm BBN} & \leq & 148.6\, \xi_T^{4}\big|_{\rm rh}\,,  \nonumber \\
g_{*}^{\rm hid}\big|_{\rm CMB}  &\leq & 50 \, \xi_T^{4}\big|_{\rm rh}  \, . \label{eq:gstarboundMSSM} 
\end{eqnarray}

Taking the case of $SU(N)$ for concreteness, the unconfined gluons
will contribute $g_{*}^{\rm hid}|_{\rm rh} = 2(N^2-1)$ at the time
they are first produced in the early universe. This number will
persist for as long as the group remains unconfined. Taking a
democratic ansatz $\xi_T |_{\rm rh} =1$, which differs from the
previous $\xi=1$ ansatz by a mild $g_S, g_S'$ dependence, this limits
the hidden sector gauge group to $SU(8)$ or smaller for $1\,{\rm MeV}
\gappeq \Lambda \gappeq 0.1\,{\rm eV}$, and $SU(5)$ or smaller for
$\Lambda \lappeq 0.1\,{\rm eV}$. Larger unconfined groups can be
accommodated with (mild) preferential reheating favoring the visible
sector.

Axions produced at the time of glueball decay will be relativistic,
and thus contribute to the bounds in~(\ref{eq:gstarbound}) as
well. For simplicity, let us consider the case in which the lifetime
of the intermediate glueball state is brief relative to the Hubble
parameter at that epoch. As argued in the previous section, this will
be true of most cases in which the glueball decays before the onset of
BBN. Under these circumstances we may approximate the thermodynamics
by assuming all of the entropy in relativistic gluons is transmitted
into a \textit{single} species of relativistic axions.

As a result, the resulting fluid of relativistic axions will
experience a `reheating' proportional to $(g_{*s}^{\rm hid})^{1/3}$,
partially off-setting the visible sector reheating associated with the
factors in~(\ref{eq:xiBBN}) and~(\ref{eq:xiCMB}). Again assuming the
reheat temperature is such as to populate the entire MSSM~field
content, the bounds in~(\ref{eq:gstarboundMSSM}) are modified to
\begin{eqnarray} g_{*}^{\rm hid}\big|_{\rm BBN} & \leq & \frac{148.6}{(g_{*s}^{\rm hid} (T_{\rm rh}))^{4/3}} \, \xi_T^{4}\big|_{\rm rh}\,,  \nonumber \\
g_{*}^{\rm hid}\big|_{\rm CMB}  &\leq & \frac{50}{(g_{*s}^{\rm hid} (T_{\rm rh}))^{4/3}}\, \xi_T^{4}\big|_{\rm rh}  \, . \label{eq:gstarboundMSSMaxion} 
\end{eqnarray}
In the case of democratic reheating ($\xi_T |_{\rm rh} =1$) the bound
in~(\ref{eq:gstarboundMSSMaxion}) arising from CMB observations falls
below unity for $g_{*s}^{\rm hid} (T_{\rm rh}) > 18$, suggesting that
a glueball which decays into axions prior to the time of CMB formation will
produce too much dark radiation for \textit{any} hidden sector gauge
group larger than $SU(3)$. Of course, larger progenitor gauge
groups can be entertained if some preferential reheating into the
visible sector is engineered. It is worth pointing out that in a world where the visible sector consists of solely the Standard Model, the visible sector reheats less (by a factor of two) and consequently \textit{any} decay into a single axion species will violate $\Delta N_{\rm eff}$ bounds in the democratic reheating limit.

\vspace{.5cm}\noindent\textbf{Compatibility with String Model Building.}

Let us briefly turn to the question of whether these bounds are compatible with string-inspired supergravity models. The answer depends on the string theory in question as well as on the uplifting and moduli stabilization scheme. In the heterotic theories, the string scale is fixed one order of magnitude above the GUT scale. Compatibility with low-energy observations fixes the gauge coupling at the GUT scale to roughly $\alpha_{\text{GUT}}=.03$, and the gauge couplings of the other hidden sector gauge groups are then expected to be in a similar range, $\mathcal{O}(10^{-1})-\mathcal{O}(10^{-2})$. In the type II theories, the string coupling is usually taken to be $0.1<g_s<1$, where the lower bound comes from compatibility with the truncation of the string $\alpha'$ expansion. The string scale and the actual gauge coupling on the stack of $D_{3+q}$-branes depends on the volume $\mathcal{V}$ of the overall compactification manifold and the volume $\mathcal{V}_q$ of the $q$-cycle wrapped by $D_{3+q}$-branes and is given by
\begin{equation}
\alpha_q=\frac{g_s}{2 \mathcal{V}_q}\,,\qquad \frac{M_s}{M_P}=\frac{g_s}{\sqrt{4\pi \mathcal{V}}}\,.
\end{equation}
 Hence for a high string scale (around the
GUT scale) we get $\mathcal{V}=\mathcal{O}(10)-\mathcal{O}(100)$ and
consequently $\mathcal{V}_q=\mathcal{O}(1)-\mathcal{O}(10)$, which
means $\alpha_q=\mathcal{O}(10^{-1})-\mathcal{O}(10^{-2})$. For
intermediate ($\mathcal{V}\sim\mathcal{O}(10^{13})$) or TeV-scale
($\mathcal{V}\sim\mathcal{O}(10^{27})$) string scales, much larger
values for $\mathcal{V}_q$ and thus much lower values for $g_q$ are
possible, and it then becomes a question of the moduli stabilization
and the uplifting scheme to answer whether the K\"ahler moduli can
be consistently stabilized in this regime.

\vspace{.5cm}\noindent\textbf{Discussion.}

\begin{figure*}[th]
  \includegraphics[width=1\columnwidth]{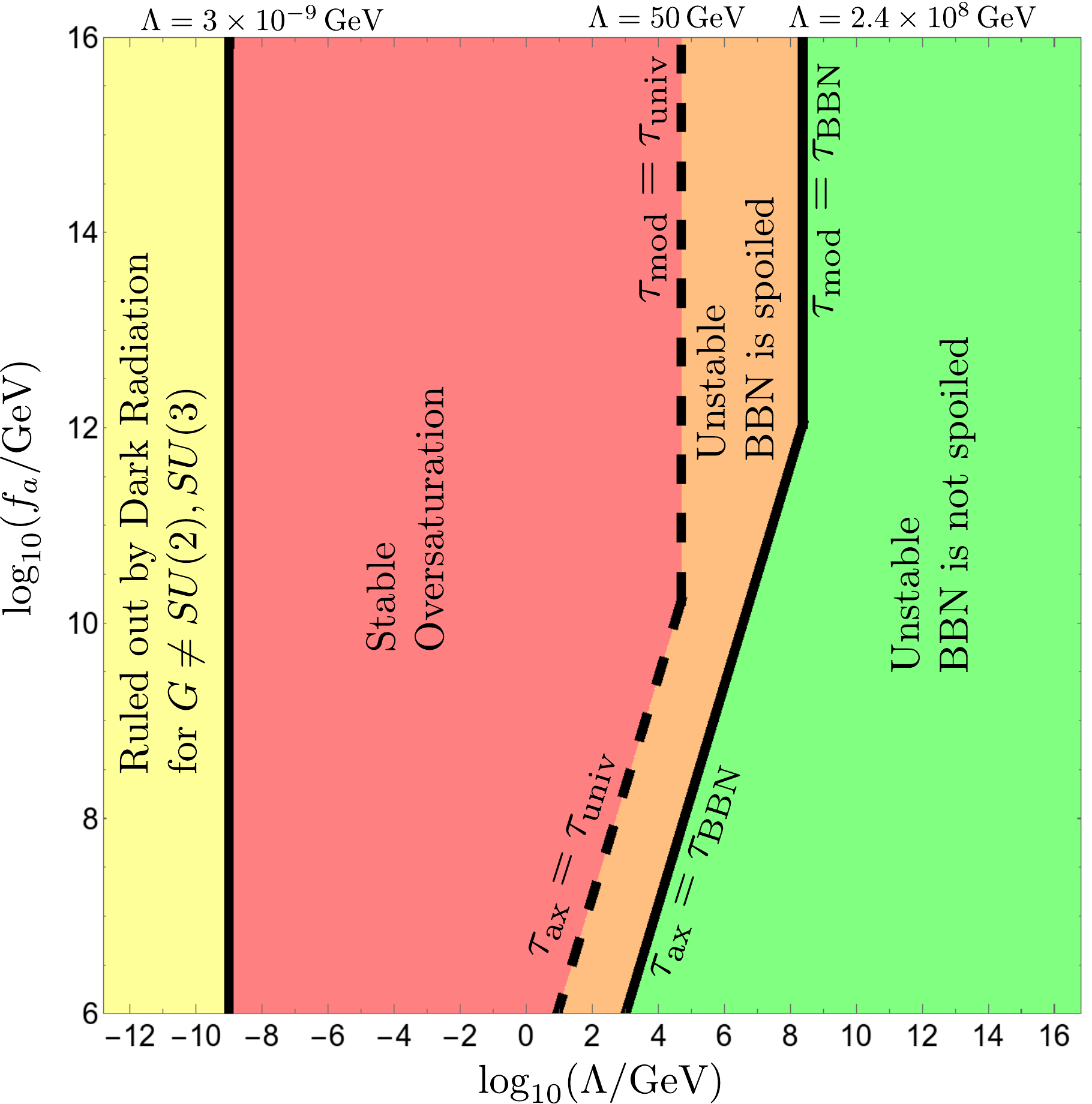} 
  \caption{Constraint regions for $\Lambda$ and $f_a$ in the democratic scenario $\xi=1$
  with the modulus decay operator suppression scale taken to be $M_s=M_{\rm GUT}$. The red and green regions are
  ruled out by a stable oversaturating glueball relic and an unstable glueball that spoils BBN; the yellow region is
  ruled out for $G\neq SU(2), SU(3)$ by dark radiation constraints; the blue region is allowed since the glueball is 
  unstable and decays to moduli or axions prior to BBN.}
  \label{fig:summaryPlots}
\end{figure*}

We have shown that theories with multiple disconnected gauge sectors --- as is typical
in string theory --- will often suffer a dark glueball problem.

In the simplest scenario of symmetric reheating via inflaton or
modulus decay, any confining Yang-Mills hidden sector with confinement
scale $\Lambda \gtrsim 3\, {\rm eV}$ will generate too much cold dark
matter to be compatible with current measurements. 
A wide range of values of the ultraviolet gauge coupling naturally give rise
 to such $\Lambda$ via renormalization group evolution,
leading to the overproduction of dark matter.  Thus, these
considerations can rule out any model with even one such Yang-Mills
sector, including string models. Confinement at scales below
$1\,{\rm eV}$ often does not solve the problem either, as constraints
on the number of relativistic degrees of freedom at the time of BBN
and CMB formation require hidden sectors no larger than $SU(5)$
($SU(3)$) if the visible sector is comprised of the MSSM (Standard
Model); hidden sectors motivated by string theory are frequently much
larger.

\vspace{.3cm}
The above statements depend, however, on two critical assumptions:
glueball stability and symmetric reheating. We studied whether relaxing
either of these assumptions may solve the glueball problem.

In fact, string theory provides a natural avenue for
relaxing the first assumption:
glueballs in hidden sectors might decay into light string moduli or
axions. If the glueball decay $\phi_{\rm} \to \chi \chi$ to moduli
happens before the epoch of BBN, the dark glueball problem maps onto
the more familiar cosmological moduli problem \cite{Banks:1995dp,Banks:1995dt,Coughlan:1983ci,deCarlos:1993wie,Kawasaki:1994af}. The latter is
solved if the lightest modulus is sufficiently heavy (we have taken
$m_{\chi} = 50\,{\rm TeV}$ here), suggesting that gauge groups with
$\Lambda \gappeq \mathcal{O}(100\,{\rm TeV})$ should produce an
acceptable cosmology. But a more precise treatment of the effective
field theory that governs glueball decays to moduli suggests much
smaller decay rates, such that the glueball decays to moduli prior to
BBN only if $\Lambda \gappeq 10^8 \,{\rm GeV}$. Thus, with other
assumptions fixed, seventeen orders of magnitude in the confinement scale
($3\, {\rm eV} \lesssim \Lambda \lesssim 10^{17}\, {\rm eV}$) remains
afflicted by the dark glueball problem.

Alternatively the glueball could decay into axions, which are
generically present in string constructions. If the axion has a mass
$m_a \ll m_{\rm gb}$ the glueball decays into a relativistic species
that is subject to $\Delta N_{\rm eff}$ bounds. In the simplest case
of a single relevant axion there are a very large number
of degrees of freedom  transferring entropy to a single field. This
tends to make the $\Delta N_{\rm eff}$ constraints more severe,
restricting the hidden sector to be no larger than $SU(3)$ in the case
when the visible sector is comprised of the MSSM (no larger than
$SU(2)$ if the visible sector is solely the Standard Model). 

There are other bounds on glueball decays into axions: if the glueball
does not decay into moduli before BBN commences, then it must decay
into axions before BBN commences, otherwise the glueball spoils BBN in
the same manner as in the cosmological moduli problem.  Since axions
can only appear in the low-energy effective Lagrangian through
derivative interactions, a decay operator consistent with the
underlying theory must be suppressed by the factor
$(\Lambda/f_a)^8$. If one takes $f_a \simeq M_s\simeq 10^{16}\, {\rm
  GeV}$, as is typical in string theory, then none of the parameter
space that isn't rescued by decay to moduli ($\Lambda \lappeq 10^8
\,{\rm GeV}$) can be rescued by glueball decay to axions, in which
case the glueball still dominates the energy density of the universe
at the time of BBN.

These results apply in the democratic reheating scenario $\xi=1$. For 
for a modulus decay operator suppression scale $M_s=M_{\rm GUT}$, the 
combined constraints from the glueball relic abundance, dark radiation, and nucleosynthesis
are taken into account in Figure \ref{fig:summaryPlots}.
We emphasize that our results hold whether or not we assume
low-scale supersymmetry, with only mild changes in the numerical
values quoted here. Note that in Figure \ref{fig:summaryPlots}
about two thirds of the parameter space is ruled out. 

We would like to emphasize one of our major points. For one Yang-Mills
sector the ruled out parameter space in Figure \ref{fig:summaryPlots}
is only somewhat constraining. However, \emph{many} hidden Yang-Mills
sectors are typical in string theory, and in such models each
glueball must fall within the allowed window. This is increasingly
difficult to achieve as the number of hidden Yang-Mills sectors
increases if the ultraviolet parameters are relatively well
distributed by the physics of moduli stabilization. Thus, though only
two thirds of the parameter space are ruled out for one Yang-Mills sector,
the constraints are much more stringent for many such sectors.

\vspace{.2cm} 
We believe that this strongly motivates the study of asymmetric reheating models \cite{Adshead:2016xxj}
in string theory, i.e.\ violating the second of our assumptions. In
this case an
initial condition, presumably through the dynamics of inflaton decay
(or the decay of other scalar fields), ensures that the bulk of the
reheating occurs in the visible sector.  This may be quantified by a
large visible to dark sector entropy density ratio $\xi=s/s'$ at the time of
reheating. 

Asymmetric reheating may aid the situation by relaxing $\Delta N_{\rm eff}$
constraints or by lowering the glueball relic abundance.  For the case of very low scale confinement ($\Lambda \ll
1\,{\rm eV}$) the bounds from $\Delta N_{\rm eff}$ considerations are
capable of accommodating larger rank hidden groups with a relatively
mild preference for the visible sector --- though the preference must
become ever more acute if there are multiple such hidden sectors. By
contrast, evading overproduction of glueball dark matter requires
$\xi \gg 1$ in large regions of the realistic UV parameter
space, often as large as $\xi = 10^{10}\,-\,10^{20}$. Such a large asymmetric reheating is 
much larger than typically necessary to avoid dark radiation problems.

\vspace{.3cm} 
Of course, if a hidden sector is
accompanied by gauge-charged matter that gives rise to baryon or meson states with masses below that of the glueball, then it may be
possible (though not immediate) to avoid the dark glueball problem. But note that these fortuitous outcomes must
arise for \textit{all} problematic hidden sectors for the construction
to be cosmologically viable. 

We therefore conclude that there is ample opportunity for a typical
string construction to experience a cosmologically fatal dark glueball
problem.  Our results motivate further research into the assurance of
high confinement scales by moduli stabilization, and also the thorny issue of reheating, with
emphasis on establishing a strong preference for the visible sector.

\vspace{.2cm}
\noindent{\bf Acknowledgments.} 
We would like to thank Pran Nath and Matt Reece
for useful discussions. The work of JH is supported by
NSF grant PHY-1620526. BDN is supported by NSF grant PHY-1314774.
The work of FR is supported by the German Science Foundation 
(DFG) within the Collaborative Research Center (SFB) 676 
``Particles, Strings and the Early Universe''.

\bibliography{refs}

\end{document}